\documentstyle{article}
\newcommand{\n}{\nonumber}
\newcommand{\bea}{\begin{eqnarray}}
\newcommand{\eea}{\end{eqnarray}}
\newcommand{\be}{\begin{equation}}
\newcommand{\ee}{\end{equation}}
\newcommand{\ph}{\phi}
\newcommand{\fr}{\frac}
\newcommand{\tl}{\tilde}
\newcommand{\sm}{\sum_n}
\newcommand{\ld}{\ldots}
\newcommand{\gm}{\Gamma}
\newcommand{\sq}{\sqrt}
\newcommand{\md}{\mid}
\newcommand{\al}{\alpha}

\title{Coherent States\\ of\\ Non-Linear Lie algebras:\\ Applications in Quantum Optics.}
\author{V. Sunilkumar, B. A. Bambah$^1$\thanks{e-mail:
bindu@panjabuniv.chd.nic.in},R. Jagannathan$^2 $\thanks{e-mail jagan@imsc.ernet.in}
P. K. Panigrahi\thanks{panisp@uohyd.ernet.in} and V.
Srinivasan\thanks{vssp@uohyd.ernet.in}\\
School of Physics,
University of Hyderabad,
Hyderabad, Andhra Pradesh,
500 046, India,\\
     $^1$Centre for Advanced Study in Mathematics,
Panjab University,
Chandigarh,
160 014, India, \\
and
$^2$Institute of Mathematical Sciences, Taramani, Chennai, India}

\begin{document}
\maketitle

\begin{abstract}
We present a general unified approach for finding the coherent states of 
 polynomially deformed algebras such as the quadratic and Higgs algebras,
which are relevant
for various multiphoton processes in quantum optics.
We give a general procedure to map
these deformed algebras to appropriate Lie algebras.
This is used , for the non compact cases, to obtain the annihilation 
operator coherent states,
 by finding the canonical
conjugates of these operators.
 Generalized coherent states, in the
Perelomov sense also  follow from this construction.
This allows us to explicitly construct coherent states associated with 
various quantum optical systems.
\end{abstract}
\section{Introduction}

Till recently , in quantum optics, only linear Lie Algebras 
have been used to give mutiphoton coherent (including squeezed) states.
However in other fields of physics, particularly string theories,
both infinite dimensional algebras as well as non-linear
"dynamical algebras" have been utilised for highly  non-linear systems .
Notable among these are the infinite dimensional Kac Moody and W Algebras and "q-algebras"
and the finite dimesional Polynomial algebras such as the quadratic and "Higgs " algebras \cite{drin,tj}.
The quadratic algebra was discovered by Sklyanin \cite{skl1,skl}, in the context
of statistical physics and field theory. The well-known Higgs algebra, a
cubic algebra, was manifest in the study of the dynamical symmetries of the
quantum oscillator and the Coulomb problem in a space of constant
curvature\cite{le,higgs}.
These algebras have now found a place in quantum optics with the observation that quantum optical Hamiltonians  
describing multiphoton processes have  symmetries which can be described by polynomially deformed
SU(1,1) and SU(2) algebras \cite{ze1,kara}. In particular , as we shall see later, the trilinear boson Hamiltonian
associated with 
various nonlinear processes in quantum optics, such as parametric
amplification, frequency conversion, Raman and Brillouin scattering, 
and the interaction of two-level atoms with a single-mode radiation 
field, has the quadratically deformed SU(1,1) algebra as a dynamical symmetry. \cite{brif}.
Similarly  the symmetry algebra for the quadratic boson Hamiltonian for four photon processes is the Higgs algebra.
By this we mean that the Hamiltonian can be written as
$H=aN_0+bN_-+cN_+$ with the N's satisfying a quadratic algebra or a Higgs Algebra.

In this paper, we present a unified approach for finding the
coherent states (CS) of these  algebras \cite {kla, perv}. 
Apart from its application to quantum optics, the method of construction
presented here is quite general will greatly facilitate the physical applications of
these algebras to many quantum mechanical problems.
 The construction of the CS for the non-compact cases,
is a two
step procedure. First, we find the canonical conjugates of these operators.
The CS, corresponding to the deformed algebras are then obtained by
the action of the exponential of the respective conjugate operators on the
vacuum \cite {pani} \cite{shanta} \cite{sc}; this is in complete parallel to the harmonic oscillator
case. Another CS, which in a sense to be made precise in the text, is dual
to the first one, naturally follows from the above construction. We also
provide a mapping between the deformed algebras  and their undeformed
counterparts. This connection is then utilized to find the CS in the
Perelomov sense\cite{perv}. Apart from obtaining the known CS of the
$SU(1,1)$ algebra, we construct the CS for the quadratic, cubic and
higher order polynomial algebras. Although our method is general, we will confine
ourselves here to finding the CS of the deformed $SU(1,1)$ and $SU(2)$
algebras.\\
\section{Polynomial Algebras as Symmetries of Multiphoton Hamiltonians.}
A Polynomial deformation of a Lie algebra is defined in the following fashion in the
Cartan-Weyl basis,
\be
[H\, , \,E_{\pm}]  = \pm E_{\pm}\,\,\,\,\,, \,\,\,\, [E_+\, , \,E_-]  =  f(H)\,\,\,\,,     
\ee
where $f(H)$ is a polynomial function of H.
The corresponding Casimir can be written in the form \cite{rocek}, 
\bea
C&=&E_- E_+ \,+\,g(H)\,\,\,, \n\\
 &=& E_+E_-\,+\,g(H-1)\,\,\,.
\eea
Here, $f(H)=g(H)-g(H-1),\,\,g(H)$ can be determined up to the addition of
a constant. The eigenstates are characterized by the values of the Casimir
operator and the Cartan
subalgebra $H$.

In particular, a polynomial deformation of $Sl(2,R)$ is of the form
$N_0=j_0, N_+=F(j_0,j)j_+, N_-=F(j_0,j)j_- $ where the $j_i$ are the ordinary $Sl(2,R)$ generators.
$[N_0,N_{\pm}]=\pm N_{\pm}$ and $[N_+,N_-]=F(N_0)$ \cite{ rocek,curt}.
 When $F(N_0)$ is quadratic in $N_0$ the algebra is called quadratic algebra
and if it is cubic in $N_0$  the "Higgs" algebra results.

 When we consider Hamiltonians 
describing multiphoton processes  it is not difficult to see that the symmetries of such Hamiltonians form
non-linear Lie algebras.
For example consider the 
triboson Hamiltonian, 
\begin{equation}
H=w_1a^{\dagger}_1a_1+
w_2a^{\dagger}_2a_2+
w_3a^{\dagger}_3a_3+
\kappa a_1a_2^{\dagger}a_3^{\dagger}+
\kappa a_1^{\dagger}a_2a_3
\end{equation}
Defining
\bea
N_0&=&(-a^{\dagger}_1a_1+
a^{\dagger}_2a_2+
a^{\dagger}_3a_3)/2    \nonumber \\                
N_+&=& a_1a_2^{\dagger}a_3^{\dagger} \nonumber \\             
N_-&=& a_1^{\dagger}a_2a_3                                 
\eea
one can show that
$N_+,N_-,N_0$ satisfy a quadratic algebra.

Another example is that of the  anharmonic oscillator with the Hamiltonian given by:
\be
H=\frac{1}{m}(a^{\dagger}a+\frac{1}{2})+
\frac{1}{n}(b^{\dagger}b+\frac{1}{2})
\ee
The operators
$N_+=a^m(b^{\dagger})^n$,
$N_-=b^n(a^{\dagger})^n$,
$N_0=\frac{1}{m}(a^{\dagger}a+\frac{1}{2})-
\frac{1}{n}(b^{\dagger}b+\frac{1}{2})$\\
commute with H, and form a quadratic algebra for m=1, n=2 with:\\
$f(H)=-3N_0^2+2HN_0+H^2-3/4$.

It can be brought to a more recognizable form
after performing an invertible basis transformation :\\
$J_+=\frac{1}{\sqrt{3}}N_+$,
$J_-=-\frac{1}{\sqrt{3}}N_-$, 
$J_0=N_0-H/3$,
where
$J_+,J_-,J_0$ satisfy quadratic algebra with
$f(J_0)=J_0^2+(\frac{1}{4}-\frac{4}{9}H^2)$ .\\
Thus the quadratic algebra represents the larger dynamical symmetry group of the anisotropic
Harmonic oscillator.          

For general  multiphoton Hamiltonians:
\begin{equation}
H=\sum_{i=0}^{1}w_ia^{\dagger}_ia_i+\kappa(a_0)^m(a_1^{\dagger})^n+c.c
\end{equation}
we can define $N_0,N_-,N_+$ in an analogous way\\
\bea
N_+&=&a_0^m(a_1^{\dagger})^n \nonumber\\
N_-&=&a_1^n(a_0^{\dagger})^m \ nonumber\\
N_0&=&\frac{1}{m+n}(a_1^{\dagger}a_1-a_0^{\dagger}a_0)
\eea
and show that we get n-dimensional Polynomial algebras as symmetry algebras for these Hamiltonans.

Similarly  n-photon Dicke Models with Hamiltonians of the form:
\begin{equation}
H=\sum{_i}^{n}\sigma_0\epsilon + w_1a^{\dagger}_1a_1+
\kappa\sum_i\sigma_-(i)(a_1^{\dagger})^n
\kappa\sum_i\sigma_+(i)(a_1)^n 
\end{equation}
can be written as:
\be
H=aN_0+bN_+ +cN_-.
\ee
with
\begin{eqnarray}
N_0=\sum{_i}^{n}\sigma_0 + a^{\dagger}_1a_1  \nonumber \\
N_-=\sum_i\sigma_-(i)(a_1^{\dagger})^n  \nonumber \\
N_+=\sum_i\sigma_+(i)(a_1)^n 
\end{eqnarray}
 satisfying  a polynomial Lie Algebra of order n.

\section{Construction of Coherent States of Non-Linear Lie Algebras:Formalism.}
 Having seen that polynomially deformed Lie algebras occur in a large class of systems in
 quantum mechanics and quantum optics in particular, we now give the formalism  
for the 
 construction of coherent states of these algebras. We do so by an "Inonu-Wigner" type for construction
of finding canonical conjugate of the annihilation operator and find the eigenstate of the annihilation operator
 by  acting the exponential of the conjugate operator on the vacuum.
This
approach is then extended to the deformed algebras in a straightforward way. 
In the next section we shall show how this fromalism can used to explicitly construct the coherent states
for application to mutiphoton processes.

First introduce our method by first considering SU(1,1) for which the generators satisfy the commutation relations
\be
   [K_+\, , \,K_-]  =  -2K_0  
\,\,,\,\,[K_{0} \,,\,K_{\pm}] =  \pm K_{\pm}\,\,\,\,.
\ee
Thus for this case
one finds, $f(K_0)= -2K_0$ and $g(K_0)=-K_0 (K_0 + 1)$. The quadratic
Casimir operator is given by $C=K_- K_+ + g(K_0) =K_-K_+ - K_0(K_0+1)$.
$\tilde{K_{+}}$, the canonical conjugate of $K_- $, satisfying
\be
[K_- \,,\,\tilde{K_+}]=1\,\,\,\,,
\ee
can be written in the form,
\be
\tilde{K_+}=K_+ F(C,K_0)\,\,\,\,.
\ee
Eq.(11) then yields,
\be
F(C,K_0)K_-K_+ - F(C,K_0-1)K_+K_- = 1\,\,\,\,;
\ee
making use of the Casimir operator relation given earlier, one can solve
for $F(C,K_0)$ in the form,
\be
F(C,K_0)=\frac{K_0 + \alpha}{C+K_0(K_0 +1)}\,\,\,\,.
\ee
The constant, arbitrary, parameter $\alpha$ in F can be determined by
demanding that Eq.$(11)$ is valid in the entire Hilbert space. For the
purpose of clarity, we illustrate this point, with the one oscillator
realization of the $SU(1,1)$ generators.

The ground states defined by $K_- \mid 0>=\frac{1}{2} a^2 \mid 0>=0$, are, 
  $\mid 0>$ and $\mid 1>=a^{\dag}\mid 0>$, in terms of the oscillator
Fock space. 
Making use of the results,
\be
K_0 \mid 0> =\frac{1}{4}(2a^{\dagger}a + 1) \mid 0> =\frac{1}{4} \mid 0>\,\,\,\,,
\ee
and
\be
C \mid 0>=\frac{3}{16}\mid 0>\,\,\,\,,
\ee
we find that, $[K_-\,,\,\tilde{K_+}]\mid 0>=K_-\tilde{K_+}\mid 0>$, yields $\alpha =\frac{3}{4}$. Similarly, for the other case
\be
[K_-\,,\,\tilde{K_+}] \mid 1>=\mid 1>\,\,\,\,,
\ee
leads to $\alpha = \frac{1}{4}$.
Hence, there are two disjoint sectors characterized by the $\alpha$ values
$\frac{3}{4}$ and $\frac{1}{4}$, respectively. These results match
identically with the earlier known ones \cite{shanta}, once we rewrite
$F$ as,
\be
F(C,K_0)= \frac{K_0 + \alpha}{C + K_0(K_0 + 1)}\,\,\,\,,
\ee
\be
 = \frac{K_0 + \alpha}{K_- K_+}\,\,\,\,.
\ee
The unnormalized coherent state $\mid\beta>$, which is the annihilation
operator eigenstate, i.e, $K_-\mid \beta >=\beta \mid \beta >$, is given
in the vacuum sector by
\be
\mid \beta> = e^{\beta \tilde{K^+}} \mid 0>\,\,\,\,.
\ee
Analogous construction holds in the other sector, where $\alpha=\frac{1}{4}$.
These states, which provide a realization of the Cat states\cite{hill},  play a
prominent role in  quantum measurement theory. As has
been noticed earlier\cite{shanta},  $[K_- \,,\,\tilde{K_+}]=1$, also yields,
\be
[\tilde{K_{+}^{\dagger}}\,,\,K_+]=1\,\,\,\,.
\ee
>From this, one can find the eigenstate of $\tilde{K^{\dagger}_{+}}$
operator, in the form,
\be
\mid \gamma>=e^{\gamma K_+}\mid 0>\,\,\,\,.
\ee
This CS, after proper normalization is the well-known Yuen state\cite{yuen}.
Our construction can be easily generalized to various other realizations
of the $SU(1,1)$ algebra. \\

We now extend the above procedure to the quadratic algebra which is the relevant algebra in considering the
coherent state of trilinear boson Hamiltonians \cite{junk}.
A example of such an algebra is the following:
\be
[N_0\,,\,N_{\pm}]=\pm N_{\pm}\,\,\,,\,\,\,[N_+ \,,\,N_-]=\pm 2N_0 + a N_{0}^{2}\,\,\,\,.
\ee
where the positive sign of $2N_0$ indicates a polynomially deformed SU(2) and a negative sign indicates polynomially deformed
SU(1,1).
In this case, $f_1(N_0)=\pm 2N_0 + a N_{0}^2=g_1(N_0)-g_1(N_0 -1)$, where,
\be
g_1(N_0)= \pm N_0(N_0 + 1)+\frac{a}{3}N_0(N_0+1)(N_0 + \frac{1}{2})\,\,\,\,. 
\ee
Representation theory of the quadratic algebra has been studied in the
literature\cite{rocek}; it shows a rich structure depending on the values
of `a'. In the non-compact case, i.e, for the values of `a' such that the
unitary irreducible representations (UIREP) are either bounded below or
above, we can construct the canonical conjugate $\tilde{N_{+}}$ of $N_-$
such that $[N_-\,,\,\tilde{N_+}]=1$. It is given
by $\tilde{N_+}=N_+F_{1}(C,N_0)$, with
\be
F_{1}(C,N_0)=\frac{N_0 + \delta}{C-N_0(N_0 +1)-\frac{a}{3} N_0(N_0+1)(N_0+\frac{1}{2})}\,\,\,\,.
\ee
As can be easily seen, in the case of the finite dimensional UIREP,
$\tilde{N_+}$ is not well defined since $F_1(C,N_0)$ diverges on the
highest state. As mentioned earlier, the values of $\delta$ can be fixed
by demanding that the relation, $[N_-\,,\,\tilde{N_+}]=1$, holds in the
vacuum sector $\mid v >_i$ where, $\mid v >_i$ are annihilated by $N_-$.
This gives $N_-\tilde{N_+}\mid v>_i=\mid v>_i$, which leads to
$(N_0+\delta)\mid v>_i=\mid v>_i$, the value of the Casimir operator,
$C=N_-N_+ + g_1(N_0)$, can be easily calculated.
Hence, the unnormalized coherent state $\mid \mu>_i: N_-\mid \mu>_i=
 \mu \mid \mu>_i$ is given by $e^{\mu \tilde{N_+}}\mid v>_i$.
The other coherent state originating from
$[\tilde{N_{+}^{\dagger}}\,,\,N_+]=1$ is given by $\mid \nu>_i=e^{\nu N_+}\mid v>_i$. This can be
recognized as the (unnormalized) CS in the Perelomov sense. Depending on the
UIREP being infinite or finite dimensional, this quadratic algebra can
also be mapped in to $SU(1,1)$ and $SU(2)$ algebras, respectively; leaving
aside the commutators not affected by this mapping, one gets,
\be
[N_+\,,\,\bar{N_-}] =- 2bN_0\,\,\,\,;
\ee
where $b=1$ corresponds to the $SU(1,1)$ and $b=-1$ gives the $SU(2)$ algebra.
Explicitly,
\be
\bar{N_-} = N_- G_1(C,N_0)\,\,\,\,,
\ee
and
\be
G_1(C,N_0) = \frac{(N_{0}^2 - N_0)b+\epsilon}{C-g_1(N_0 - 1)}\,\,\,\,,
\ee
$\epsilon$ being an arbitrary constant.
One can immediately construct CS in the Perelomov sense as $U\mid v>_i$, where
$U=e^{\xi N_+ -\xi^{\ast}N_-}$.
For the compact case, the CS are analogous to the spin and atomic coherent
states\cite{spin,atcs}.

The Cubic algebra, which is also popularly known as the Higgs algebra in the
literature, manifested in the study of the degeneracy structure of
eigenvalue problems in a curved space\cite{higgs}.
The generators satisfy,
\be
[M_{0}\,,\,M_{\pm}] = \pm M_{\pm}\,\,\,,\,\,[M_{+}\,,\,M_{-}] = 2cM_0\,+\,4hM_{0}^3\,\,\,\,,
\ee
where, $f_2(M_0) = 2cM_0\,+\,4hM_{0}^3 = g_2(M_0)-g_2(M_0 - 1)$, and
\be
g_2(M_0) = cM_0(M_0 + 1)\,+\,hM_{0}^2(M_0 + 1)^2\,\,\,\,.
\ee
Analysis of its representation theory yields a variety of UIREP's, both finite
and infinite dimensional, depending on the values of the parameters
$c$ and $h$ \cite{zed}. Physically, $h$ represents the curvature of the
manifold. In the non-compact case the canonical conjugate is given by,
\be
\tilde{M_+} = M_+ F_{2}(C,M_0)\,\,\,\,,
\ee
where,
\be
F_{2}(C,M_0) = \frac{M_0 + \zeta}{C-cM_0(M_0+1)-hM_{0}^2(M_0 + 1)^2}\,\,\,\,.
\ee
As before, the annihilation operator eigenstate is given by
\be
\mid \rho>_i= e^{\rho\tilde{M_+}}\mid p>_i\,\,\,\,,
\ee
where, $\mid p>_i$ are the states annihilated by $M_-$. Like the previous
cases, the dual algebra yields another coherent state.  
This algebra can also be mapped in to $SU(1,1)$ and $SU(2)$ algebras, as
has been done for the quadratic case:
\be
[M_+\,,\,\bar{M_-}]=-2dM_0\,\,\,\,,
\ee
where, $d=1$ and $d=-1$ correspond to the $SU(1,1)$ and $SU(2)$ algebras respectively.
Here,
\be
\bar{M_-}=M_-G_2(C,M_0)\,\,\,\,,
\ee
where,
\be
G_2(C,M_0)=\frac{(M_{0}^2-M_0)d+\sigma}{C-g_2(M_0-1)}\,\,\,\,,
\ee
$\sigma$ being a constant.
The coherent state in the Perelomov sense is then $U\mid v>_i$, where,
$U=e^{\phi M_+ -\phi^{\ast}M_-}$.
We would like to point out that, earlier the generators of the deformed
algebra have been written in terms of the undeformed ones\cite{rocek}.
However, in our approach the undeformed $SU(1,1)$ and $SU(2)$ generators
are constructed from the deformed generators \cite{curt}.

\section{Explicit construction of the Coherent States for Physical Application.}
We now construct the state explicitly for purposes of application.
First we  show that this method , indeed, gives us well known SU(1,1) Barut-Giradello (pair coherent) states for SU(1,1) in the familair form \cite{barut}.
The action on Hilbert Space of the generators is given in the original Barut Giradello representation by:
\be
K_0 \md \ph >=(-\ph + m)\, \md \ph,m > \\
\ee
\be
K_+ \md \ph,m >= \fr{1}{\sq{2}} \sq{(m+1)(-2\ph + m)}\, \md \ph,m+1> \\
\ee
\be
K_-  \md \ph,m>= \fr{1}{\sq{2}} \sq{m( -2 \ph + m-1)}\, \md \ph m-1> \\
\ee
The Coherent state $ \md \al > $ is given by
\be
\md \al >=e^{\al \tilde{K_+}} \md \ph,0> \\
\ee
where $ [K_- \,,\,\tilde{K_+}]=1 $ and $ \tilde{K_+} =K_+ F(C,K_0)$
\be
F(C,K_0)=\fr{1}{C-g(K_0)} =\fr{1}{C+ \fr{1}{2} K_0 (K_0 + 1)} \\
\ee
\be
\md \al >=\sm \fr{\al^n }{n!} (K_+ F(C,K_0))^n \md \ph , 0> =\sm \fr{\al^n}{n!} (K_+)^n F(C,K_0) \ld F(C,K_0 +n-1) \md \ph ,0> \\
\ee
substituting the values of $F$ we get:
\bea
|\alpha>&=&\sm \fr{\al ^n}{n!} \fr{1}{-\ph }\, \fr{1}{-\ph + \fr{1}{2}} \ld \fr{1}{-\ph +
\fr{n-1}{2}} \,\, (K_+)^n \, \md \ph ,0> \nonumber \\
&=&\sm \fr{\al^n}{n!} (2)^n \fr{\gm (-2\ph )}{\gm (-2\ph + n)} \fr{\sq{n! (-2\ph
+ n-1)! }}{(\sq{2})^n \sq{\gm (-2\ph )}} \md \ph ,n> \nonumber \\
&=&C \sq{\gm (-2\ph )} \sm (\sq{2} \al )^n \fr{1}{\sq{\gm (n+1)\, \gm (-2\ph +n)}} \md \ph ,n> 
\eea
Which is precisely the well-known state of Barut and Giradello upto the normalization coefficient C.

First we consider a quadratic algebra and then show the general explicit construction.
For the quadratic case we take an illustrative algebra  relevant to the trilinear boson cases described in section 2.
A typical algebra satisfied by the generators is given :\\
 $[N_0\,,N_+]=N_+\,\, [N_0\,N_-]=-N_-\,\,
[N_+\,N_-]=-3N_{0}^{2} +4\epsilon N_0-\epsilon^2 $
where $\epsilon$ is a constant and various values of $\epsilon$ and suitable
basis transformations gives the symmetry algebra of most trilinear Hamiltonians \cite{junk} \cite{fern}.
The action of the operators on eigenfunctions of $N_0$ is given by:
\be
N_0 \md n>=(n+\fr{1}{2} ) \md n>
\ee
\be
N_+ \md n>=\sq{(n+\fr{3}{2} -\epsilon)(n+1)(n+ \fr{1}{2} -\epsilon)} \md n+1>
\ee
\be
N_- \md n>=\sq{(n-\fr{1}{2} -\epsilon)n(n+\fr{1}{2} -\epsilon)} \md n-1 >
\ee
Here $ g(N_0)=-(N_0 -\epsilon)(N_0 +\fr{1}{2} )(N_0 +1-\epsilon) $ and
$g(N_0)-g(N_0-1)=3N_{0}^{2} -4\epsilon N_0+\epsilon^2 $

>From our construction the CS therefore is:
\be
\md \al >=e^{\al \tl{N_+}} \md 0>=\sm \fr{\al^n }{n!} (\tl{N_+})^n \md 0>
\ee
Thus:
\be
|\alpha>=\sm \fr{\al^n }{n!} (N_+ F(N_0,C))^n \md 0>
\ee
Constructiong the $F's$ from  $g(N_0)$ we get:
\bea
|\alpha>&=&\sm \fr{\al^n }{n!} (N_+)^n F(N_0)F(N_0 +1) \ld F(N_0 +n-1) \md 0> \nonumber \\
&=&\sm \fr{\al^n }{n!} (N_+)^n \fr{N_0 +\delta}{(N_0 -\epsilon)(N_0 +\fr{1}{2})
(N_0 +1-\epsilon)} \ld \fr{N_0 +n-1+\delta }{(N_0 +n-1-\epsilon)(N_0 +n-\fr{1}{2})
(N_0 +n-\epsilon)} \md 0> \nonumber \\
&=&\sm \al ^n  \fr{(-\fr{1}{2} -\epsilon)!(\fr{1}{2} -\epsilon)!}{(n-\fr{1}{2} -\epsilon)!
n!(n+\fr{1}{2} -\epsilon)!} (N_+)^n \md 0> \nonumber \\
&=&C \sq{\gm (\fr{1}{2} -\epsilon) \gm (\fr{3}{2} -\epsilon)} \sm 
\fr{\alpha^n}{\sq{\gm (n+\fr{1}{2} -\epsilon) \gm (n+1) \gm (n+\fr{3}{2} -\epsilon)}}
\md n>
\eea
$C$ is the normalization coefficient , which can be easily determined.

We now give an outline of the method of explicit construction of coherent states
for general multiphoton processes for which the generators satisfy the algebra \cite{bon}:
$[N_0\,,N_+]=N_+\,\, [N_0\,N_-]=-N_-\,\,
[N_+\,N_-]=g(N_0) -g(N_0 -1) $

The action on eigenstates of $N_0$ is given by
\be
N_0 \md j,m>=j+m \md j,m>
\ee
\be
N_+ \md j,m>=\sq{C(j)-g(j+m)} \md j,m+1>
\ee
\be
N_- \md j,m>=\sq{C(j)-g(j+m-1)} \md j,m-1>
\ee
where $ C(j)=g(j-1) $\\

Coherent state is given by
\bea
\md \al >&=&e^{\al \tl{N_+}} \md j,0> \nonumber \\
&=&\sm \fr{\al ^n }{n!} (N_+)^n \fr{N_0 +\delta }{g(j-1)-g(N_0)}
\ld \fr{N_0+n-1+\delta }{g(j-1)-g(N_0 +n-1)} \md j,0> \nonumber \\
&=&C\sm \al ^n \fr{1}{\sq{(g(j-1)-g(j)) \ld (g(j-1)-g(j+n-1))}} \md j,n>
\eea

A discussion of coherent states is incomplete without showing that these states do gice a resolution
of the identity and that they are over complete. From the resolution of the identity we have:
\be
\int d\sigma(\alpha^{*},\alpha)\, |\alpha\rangle\langle\alpha| = {\bf 1} 
\ee
Within the polar decomposition
ansatz 
\be
d\sigma(\alpha^{*},\alpha) = r \sigma(r) d\theta r dr 
\ee
with $r=|\alpha|$ and a yet unknown positive density $\sigma$  which provides the measure.
For the general case we have:
\be
2\pi\int_{0}^{\infty} dr\, \sigma(r) \, r^{2n+1} = C (g(j-1)-g(j)).....(g(j-1)-g(j+n-1))
\ee
For the various cases the substitution of the explicit value of g(j) then reduces the expression on the R.H.S
to a rational function of  Gamma Functions and the measure $\sigma$ can be found by an inverse Mellin transform.
For example for the SU(1,1) $g(j)=-\frac{1}{2}(j)(j+1)$ case the R.H.S. becomes in the BG representation
\be
2\pi\int_{0}^{\infty} dr\, \sigma(r) \, r^{2n+1} =C \frac{\Gamma(n+1) \Gamma(-2\phi+n)}{\Gamma(-2\phi)}
\ee
where C is a numerical constant
and from the inverse Mellin transform we get $\sigma(r)= C r^{-2\phi +1}K_{\frac{1}{2}+\phi}(2r)$

For the quadratic case the R.H.S. becomes
\begin{equation}
\int_{0}^{\infty} dr\, \sigma(r) \, r^{2n+1} = \Gamma(n+1)
 \, \frac{\Gamma(\frac{1}{2}-\varepsilon+n)
\Gamma(\frac{3}{2}-\varepsilon+n)}
 {\Gamma(\frac{1}{2}-\varepsilon)\Gamma(\frac{3}{2}-\varepsilon)}
 \end{equation}
 and $\sigma(r)$ can be determined to be a confluent hypergeometric function from the inverse Mellin transformation formula:

\begin{equation}
\int_0^{\infty} r^{b-1} \Phi(a,c,-r)=\frac{\Gamma(b) \Gamma(c) \Gamma(a-b)}{\Gamma(a) \Gamma(c-b)}
\end{equation}
For the general case the measure will be a Meijer's G-function.

\section{Conclusion}

To conclude, we have found a general method for constructing the coherent
states for various polynomially deformed algebras for quantum optical systems whose dynamics is governed by multilinear boson Hamiltonians.
Since the method is algebraic and
relies on the group structure
of Lie algebras, the precise nature of the non-classical behaviour
of these CS can be easily inferred from our construction. It will be of
particular interest to see the the time development of the system and
the physical role of the deformation parameters.
Since many of these  algebras are related to
quantum mechanical problems with non-quadratic, non-linear Hamiltonians,
 a detailed study of
the properties of the CS associated
with these non-linear and deformed algebras is of physical relevance \cite{vin, deb}.
This is the subject of our current and future work \cite{sun}.

The authors take the pleasure to thank Prof. S. Chaturvedi for stimulating
conversations. VSK acknowledges useful discussions with Mr. N. Gurappa.


\begin{thebibliography}{99}
\bibitem{drin} V. Drinfeld and V.Sokolov, J. Sov. Math {\bf 30} 1975 (1984).
\bibitem{tj}J. de Boer, F. Harmsze and T. Tjin,  Phys. Rep. {\bf 272},
139 (1996) and references therein.
\bibitem{skl1}E. K. Sklyanin,  Funct. Anal. Appl. {\bf 16}, 263 (1982).
\bibitem{skl}E. K. Sklyanin, Funct. Anal. Appl. {\bf 17}, 273 (1983).
\bibitem{le}M. Lakshmanan and K. Eswaran,  J. Phys. {\bf A}: Math. Gen. {\bf 8}, 1658 (1975).
\bibitem{higgs}P. W. Higgs,  J. Phys. {\bf A}: Math. Gen. {\bf 12}, 309 (1979)..
\bibitem{kara} V. P. Karassiov,  J. Phys. {\bf A}: Math. Gen. {\bf 27}, 153 (1994).
\bibitem{ze1} A. S. Zhedanov,Mod. Phys. Lett. {\bf A 7}, 507 (1992).
\bibitem{brif} C. Brif,{\sl "Coherent States for quantum systems with a trilinear boson Hamiltonian"} quant-ph/9608022  .//(The construction here is very different from ours.)
\bibitem{kla}J. R. Klauder and B-S. Skagerstam,{\sl  Coherent States}, (World
 Scientific, Singapore, 1985).
\bibitem{perv}A. M. Perelomov,{\sl  Generalized Coherent States and Their Applications}, (Springer, Berlin, 1986).
\bibitem{pani}N. Gurappa, P. K. Panigrahi and V. Srinivasan,  Mod. Phys. Lett. {\bf A 13}, 339 (1998). 
\bibitem{shanta}P. Shanta, S. Chaturvedi, V. Srinivasan, G. S. Agarwal and C. L. Mehta,  Phys. Rev. Lett. {\bf 72}, 1447 (1994).
\bibitem{sc}P. Shanta, S. Chaturvedi, V. Srinivasan and R. Jagannathan,  J. Phys.{\bf A}: Math. Gen. {\bf 27}, 6433 (1994).
\bibitem{rocek}M. Ro$\hat{c}$ek,  Phys. Lett. {\bf B255}, 554 (1991).
\bibitem{curt}T. Curtwright and C. Zachos, Phys. Lett. {\bf B243}, 237 (1990).
\bibitem{hill}M. Hillery,  Phys. Rev. {\bf A 36}, 3796 (1987); C. C. Gerry and E. E. Hach III,  Phys. Lett. {\bf A 174}, 185 (1993).
\bibitem{yuen}H. P. Yuen,  Phys. Rev. {\bf A 13}, 2226 (1976).
\bibitem{spin}J. M. Radcliffe, J. Phys. {\bf A}: Gen. Phys {\bf 4}, 313 (1971). 
\bibitem{atcs}F. T. Arecchi, E. Courtens, R. Gilmore and H. Thomas,  Phys. Rev. {\bf A 6}, 2211 (1972).
\bibitem{zed}Ya. I. Granovskii, A. S. Zhednov and I. M. Lutzenko,  J. Phys. {\bf A}: Math. Gen. {\bf 24}, 3887 (1991).
\bibitem{barut} A.O. Barut and L. Girardello, Comm. Math. Phys {\bf 21}, 45 (1971)
\bibitem{junk} F. Cannata, G. Junker and J. Trost , quant-ph/9806080,
published in {\sl Particles Fields and Gravitation} edited by J. Rembielinski,
1998, AIP Conf Proc. {\bf 453}, 209.
\bibitem{fern}D.J. Fernandez and V. Hussin, J. Phys. {bf A 32}, 3603, (1999).
\bibitem{bon}D. Bonatsos, C. Daskaloyannis and K. Kokkotas,  Phys. Rev. {\bf A 48}, R3407 (1993).  
\bibitem{vin}P. L$\acute{e}$tourneau and L. Vinet,  Ann. Phys. {\bf 243}, 144 (1995).
\bibitem{deb}J. Beckers, Y. Brihaye and N. Debergh,  "On realization of
nonlinear Lie algebras by differential operators and some physical
application'' hep-th/ 9803253.
\bibitem{sun} V. Suneel Kumar, B. Bambah, P. Panigrahi and V.
Srinivasan. {\sl "Coherent States of the Deformed Algebras} quant-ph/9905010    .
\end{thebibliography}
\end{document}